\documentclass[floats,aps,showpacs,twocolumn]{revtex4}
\usepackage{graphicx}

\newcommand{\be}{\begin{equation}}
\newcommand{\ee}{\end{equation}}
\newcommand{\ba}{\begin{eqnarray}}
\newcommand{\ea}{\end{eqnarray}}

\newcommand{\opone}{\leavevmode\hbox{\small1\kern-3.8pt\normalsize1}}

\newcommand{\e}{{\rm e}}

\begin{document}
\title{Semiclassical Theory of Chaotic Conductors}

\author{ Stefan Heusler$^1$, Sebastian M{\"u}ller$^1$, Petr Braun$^{1,2}$,
Fritz Haake$^1$}

\address{$^1$Fachbereich Physik, Universit{\"a}t Duisburg-Essen,
45117 Essen, Germany\\
$^2$Institute of Physics, Saint-Petersburg University, 198504
Saint-Petersburg, Russia}

\begin{abstract}
  We calculate the Landauer conductance through chaotic ballistic
  devices in the semiclassical limit, to all orders in the inverse
  number of scattering channels without and with a magnetic field.
  Families of pairs of entrance-to-exit trajectories contribute,
  similarly to the pairs of periodic orbits making up the small-time
  expansion of the spectral form factor of chaotic dynamics. As a clue
  to the exact result we find that close self-encounters slightly
  hinder the escape of trajectories into leads.

\end{abstract}


%73.23.-b Electronic transport in mesoscopic systems
%73.20.Fz Weak or Anderson localization (surface/interface states)
%72.20.My Galvanomagnetic and other magnetotransport effects

\maketitle

%\begin{center} Draft conductance22 of \today\end{center}
\date{\today}

\begin{figure*}
\begin{center}
  \includegraphics[scale=0.12]{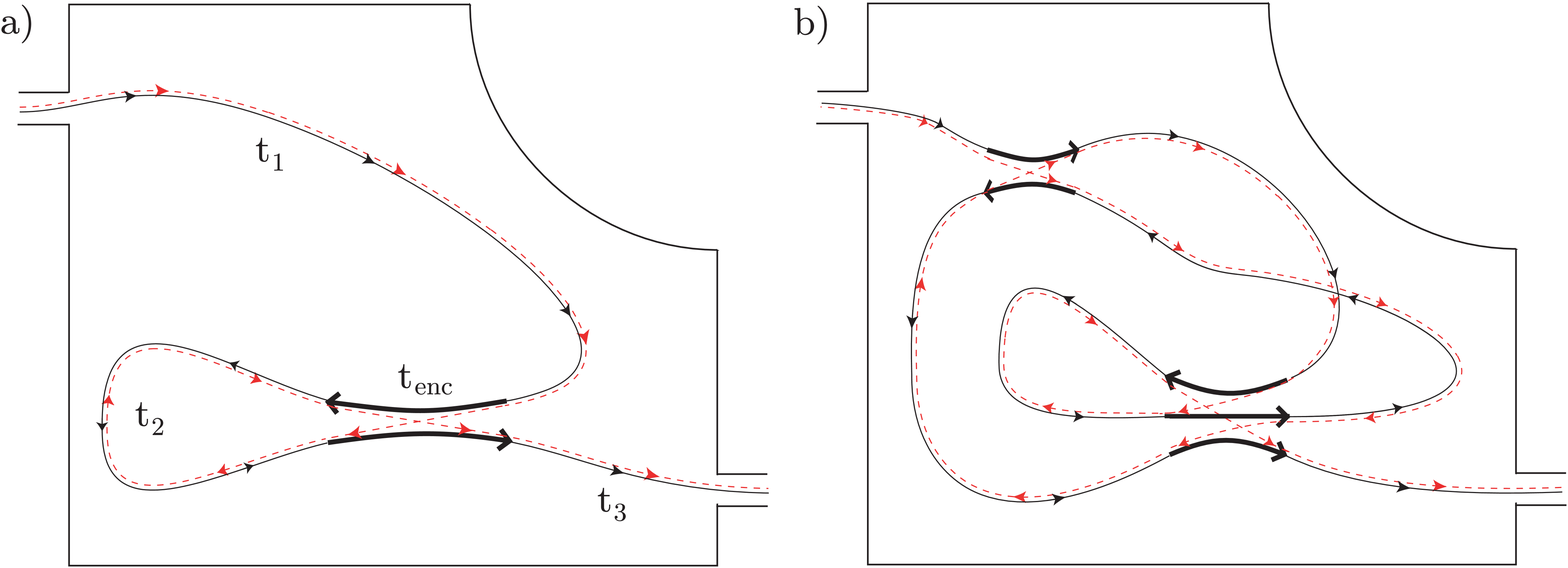}
\end{center}
\caption{Pairs of trajectories $\gamma,\gamma'$ contributing to
the conductance: a) a Richter/Sieber pair differing in a
2-encounter, b) a pair of trajectories differing in one
2-encounter and one 3-encounter. In a), $\gamma$ and $\gamma'$
differ in the sense of traversal of one  loop ($\nu=1$), and the
relative orientations of stretches in the 2-encounter lead to $\mu=0$. For b)
we have $\nu=3$; $\mu=-1$ for the 2-encounter and $\mu=1$ for the
3-encounter} \label{fig:conductance}
\end{figure*}

Electron transport through conductors is usually dominated by
scattering from disordered impurities. The pertinent theory involves
averages over random configurations of scatterers and ultimately
amounts to random-matrix theory. Typical consequences of disorder on
transport, like weak localization, actually survive for ``clean''
systems whose linear dimension is smaller than the mean free path,
provided the ballistic motion is chaotic due to irregularly shaped
boundaries of the ``cavity'' serving as the conductor
\cite{Jal00,Richhabil}. The equivalence of chaos in the ballistic
limit and impurity scattering represents a theoretical challenge
similar to that of proving the so-called Bohigas-Giannoni-Schmit
conjecture about universality of fluctuations in energy spectra of
chaotic dynamics.

Gutzwiller's semiclassical periodic-orbit theory has recently been
brought to success in explaining spectral fluctuations of individual
chaotic dynamics as faithful to the ensemble averages provided by
random-matrix theory \cite{HMBH}. Analogous attempts have been
undertaken for ballistic transport \cite{Richhabil,Richter}, employing
semiclassical techniques in Landauer's scattering approach to
conductance \cite{Landi}.  The two lowest-order terms in a
$1/N$-expansion (with $N$ the dimension of the scattering matrix) of
the conductance were thus found in agreement with RMT. Here we take up
that line of thought, with two crucial modifications to be explained
below, which yield the conductance to all orders in $1/N$. Both for
systems with and without time reversal invariance as well as for the
crossover between these two cases caused by a magnetic field we
confirm the equivalence to impurity scattering as predicted by
random-matrix theory.  Each order of the $1/N$ expansion turns out
one-to-one with the small-time expansion of the universal spectral
form factor given in \cite{HMBH}. Our confirmation of the validity of
the RMT conductance for individual chaotic cavities with arbitrary $N$
is particularly relevant for (electronic or microwave) experiments
involving a number of channels of order unity \cite{BeeHout}.

We imagine a two dimensional cavity connected to the outside by two
leads respectively supporting $N_1$ and $N_2$ channels. The
entrance-to-exit transition amplitudes for (electron) waves then form
a unitary $N\times N$ $S$-matrix where $N=N_1+N_2$.  Sub-blocks $t,t'$ of
respective size $N_1\times N_2,\,N_2\times N_1$ comprise the amplitudes for
transitions between different leads while the remaining sub-blocks
$r,r'$ refer to ``reflections'', i.~e.~transitions with entrance and
exit at the same lead. We are concerned with the conductance between
different leads, given by Landauer \cite{Landi} as $G={\rm Tr}(t^\dagger t)$,
in units of the quantum conductance $G_0=e^2/\pi\hbar$.  Random-matrix
theory yields an averaged conductance $G_\beta$, the average being over
the Wigner-Dyson circular ensembles of unitary matrices from the
orthogonal ($\beta=1$) or unitary ($\beta=2$) class \cite{RMT,
  Bee97},
\begin{equation}
\label{predic}
G_{\beta} = \frac{N_1 N_2}{N -1 + 2/\beta} =
\frac{N_1 N_2}{N}\Big\{1 +
\sum_{n=1}^{\infty}\frac{(1-2/\beta)^{n}}{N^n}\Big\}\,.
\end{equation}
We first deal with the orthogonal and unitary classes. Further
below, we shall allow for a magnetic field causing a smooth
transition between these classes. Random-matrix theory provides an
integral expression \cite{Weidi} $G(\xi)$ interpolating between
$G_1$ and $G_2$, as a parameter $\xi$ proportional to the squared
magnetic field is varied.

We shall derive a semiclassical expression for the conductance of
a single chaotic cavity. %(rather than an ensemble).
The van Vleck propagator for the amplitudes $t$ brings classical
entrance-to-exit trajectories into play \cite{Jal00,Richhabil},
$t_{nm}\propto\sum_\gamma A_\gamma\e^{{\rm i} S_\gamma/\hbar}$, with $S_\gamma$ the action and $A_\gamma$
a stability amplitude of $\gamma$; the sum includes all trajectories
connecting the leads with discrete incident angles, one for each
channel. The Landauer conductance becomes the double sum
\cite{Richhabil},
\begin{equation}\label{GSC}
{\rm Tr} (t^{\dagger} t) = \frac{1}{T_H}\,
\left\langle\sum_{\gamma, \gamma'} A_{\gamma} A_{\gamma'}^* \,
\e^{{\rm i}(S_{\gamma} - S_{\gamma'})/ \hbar}\right\rangle\,,
\end{equation}
somewhat similar to  Gutzwiller's \cite{HMBH} double sum over
periodic orbits for the spectral form factor; here $T_H$ is
the Heisenberg time, related to the volume of the energy
shell as $T_H=\Omega/ 2\pi\hbar$. The brackets $\langle\cdot\rangle$
denote an energy average over a small interval near the Fermi energy.

In the limit $\hbar\to 0$ only trajectory pairs with the action
difference $\Delta S=S_\gamma-S_{\gamma'}$ of the order $\hbar$ may
survive the averaging implied in (\ref{GSC}). The simplest such family
is the diagonal one with $\gamma=\gamma'$. To evaluate the diagonal
approximation ${\rm Tr} (t^{\dagger} t)_{\rm diag} =T_H^{-1}\langle
\sum_{\gamma = \gamma'} |A_{\gamma}|^2 \rangle$, Richter and Sieber
\cite{Richter} invoked a new sum rule which we write as
\begin{eqnarray}
 {\rm Tr} (t^{\dagger} t)_{\rm diag} &=& \frac{1}{T_H}\Big \langle \sum_{\gamma
= \gamma'} |A_{\gamma}|^2\Big \rangle =\frac{N_1 N_2}{T_H}  \int_{0}^{\infty} dT \e^{- \frac{N}{T_H}T}
 \nonumber\\\label{RSsum}
&=&
 \frac{N_1 N_2}{N}\,.
\end{eqnarray}
Like the somewhat analogous sum rule given by Hannay and Ozorio de
Almeida \cite{HOdA} for periodic orbits, the Richter-Sieber sum
reflects the ergodicity of long trajectories; in addition, it
accounts for the escape of a trajectory hitting a lead after a
dwell time  $T$ by the exponential ``survival probability''
$\e^{-\frac{N}{T_H}T}$. In spite of its quantum guise, which is
convenient for our purposes, the escape rate $N/T_H$ is, of
course, a classical quantity; Planck's constant cancels since the
number of channels is proportional to the ratio of the width of
the leads and the de Broglie wavelength. Clearly, the diagonal
approximation gives the first term of the $1/N$ expansion of the
random-matrix conductance (\ref{predic}), and even the full
conductance for the unitary symmetry class.

Higher-order terms are due to families of pairs $(\gamma,\gamma')$ of
trajectories.  The partner trajectories in a pair nearly coincide, at
least up to time reversal, in ``loops'' whose cumulative duration is
of the order $T_H$; these loops are separated by close self-encounters
of each trajectory; the encounter stretches have durations of the
order of the Ehrenfest time $T_E\propto \ln\hbar\ll T_H\propto 1/ \hbar$ and connect
the loops differently for $\gamma$ and $\gamma'$. The various families of
pairs $(\gamma,\gamma')$ can be imagined as arising from the pairs of periodic
orbits contributing to the analogous expansion of the spectral form
factor \cite{HMBH} by cutting a loop into two and pulling the two open
ends to different leads.  The cutting described actually turns the
families of pairs of periodic orbits giving the term $\tau^{n}$ in
the $\tau$ expansion of the spectral form factor into the pairs of
trajectories contributing to the $n$-th order in the $1/N$ expansion
of the conductance. The dwell time $T$ of $\gamma$ and $\gamma'$ is then the
cumulative duration of all loops and encounter stretches.

A clue to the full expansion in search is the survival probability.
That quantity is {\it a little larger than} $\e^{-\frac{N}{T_H}T}$
for all trajectory pairs except the ``trivial'' ones,
$\gamma'=\gamma$.  To see that, we focus on, say, an $l$-encounter
where $l$ stretches of $\gamma$ (and of $\gamma'$) are mutually close
at least up to time reversal and separated from the leads by loops;
each of the $l$ intra-encounter stretches has the duration $t_{\rm
  enc}$.  Survival of $\gamma$ inside the cavity during the first of
the $l$ stretches implies survival during the $l-1$ other stretches.
The whole encounter therefore does not contribute $lt_{\rm enc}$ but
only $t_{\rm enc}$ to the effective exposure time in the exponential
survival probability.  The cumulative exposure time is the sum of all
loop durations, augmented by one instead of $l$ contributions $t_{\rm
  enc}$ for each $l$-encounter, $T_{\rm exp}=\sum_{\rm loops}t_{\rm
  loop}+\sum_{\alpha}t_{\rm enc
}^\alpha=T-\sum_{\alpha}(l_\alpha-1)t_{\rm enc}^\alpha$, the index
$\alpha$ labeling the encounters.  Due to $T_E\ll T_H$ the survival
probability $\e^{-\frac{N}{T_H}T_{\rm exp}}$ differs only little from
the naively expected value $\e^{- \frac{N}{T_H}T}$, but the difference
will presently be seen to be decisively important.

We illustrate our procedure for the first correction to the
diagonal approximation which involves partner trajectories
differing in a single 2-encounter.
Fig.~1a depicts the two encounter
stretches and three loops with durations $t_i, i=1,2,3$; the
dwell time  is $T=t_1+t_2+t_3+2t_{\rm enc}$. The two encounter
stretches must be nearly mutually time
reversed since otherwise the partner trajectory would decompose
such that one of the loops would connect the two leads while the
other would close in on itself to form a periodic orbit; this is
why the first off-diagonal correction %to the diagonal approximation
requires time reversal invariance.

We may search for 2-encounters drawing a Poincar{\'e} section
$\cal P$ transverse to $\gamma$. If the section hits a
2-encounter, $\gamma$ must pierce through $\cal P$ in two points
which are almost mutually time reversed. We denote the stable and
unstable shifts of the second (time reversed) piercing relative to
the first by $s,u$. For the quantitative definition of
2-encounters we require $|u|\leq c\,,|s|\leq c$ with $c$ a
constant independent of $\hbar$ (whose exact value is irrelevant).
The coordinates $u,s$ determine the encounter duration $t_{\rm
enc}=(1/\lambda)\ln c^2/|su|$ and the action difference $\Delta
S=S_\gamma-S_{\gamma'}=us$; here $\lambda$ is the Lyapunov
constant.

The ergodic probability of the second piercing occurring with
given $s,u$ and duration $t_2$ of the loop separating the
encounter stretches is $ds\,du\,dt_2/\Omega$. We can write the
number of trajectory pairs generated by 2-encounters with the
given action difference $\Delta S$ as $d\Delta S
\int_{|s|<c,|u|<c}ds\;du\;w_T(u,s)\delta(\Delta S-us)$. The weight
function $w_T(u,s)$ is obtained from the ergodic return density
$1/\Omega$ by (i) integrating over all durations $t_2$ of the loop
separating the encounter stretches; (ii) integrating over all
positions of the Poincar{\'e} section which is equivalent to
integrating over the duration of the first loop $t_1$; (iii)
dividing by the encounter duration $t_{\rm enc}$ (the probability
to cross an encounter is proportional to its duration; in order to
obtain the number of encounters from the crossing probability the
encounter duration has to be divided out). Requiring all loop
times to be positive we get
\begin{eqnarray}\label{defw}   \nonumber
w_T(u,s) &=&
\int\frac{dt_1dt_2}{\Omega t_{\rm enc}}
\Theta(t_1)\Theta(t_2)\Theta(T-t_1-t_2-2t_{\rm enc})
 \\
&=& \frac{\left(T-2t_{\rm enc}(u,s)\right)^2}{2t_{\rm enc}(u,s)\Omega}\;.
\end{eqnarray}
The first off-diagonal term in the conductance thus reads
\begin{eqnarray}
G^{(1)}=\left\langle\int_{|s|<c,|u|<c}\!\!\!\!dsdu\, {\rm e}^{{\rm i} su/\hbar}
\sum_\gamma \frac{|A_\gamma|^2 }{T_H}w_T(s,u) \right\rangle\,.
\end{eqnarray}
The sum over $\gamma$ is now calculated using the rule (\ref{RSsum}) with
the crucial modification pointed out above, i.e. the survival
probability ${\rm e}^{-\frac{N}{T_H}(t_1+t_2+t_3+t_{\rm enc})}$. The
integrations over $t_1,t_2$ (in $w_T$) and over the dwell time $T$ can
be replaced by  integrations over all three loop durations.
The lower limit of a loop duration is {\it zero} which expresses a
second crucial ingredient of our approach: {\it the encounter
  stretches may neither overlap nor stick out of the cavity openings}
\cite{footRS}; the upper limit is infinity. The result
\begin{eqnarray}
G^{(1)}&=&\frac{N_1N_2}{T_H} \left(\prod_{i=1}^3\int_{0}^\infty dt_i
\e^{-\frac{N}{T_H}t_i}\right)   \\ \nonumber
&&\times\left<\int_{|s|<c,|u|<c}\!\!\!\!dsdu\frac{\e^{-
\frac{N}{T_H}t_{\rm enc}+{\rm i} s u/\hbar}}{\Omega t_{\rm enc}}\right>
\to - \frac{N_1N_2}{N^2}
\end{eqnarray}
obtains a factor $T_H/N$ from each loop-time integration. To calculate
the integral over $s,u$ we expanded in powers of the encounter
duration, $\e^{-\frac{N}{T_H}t_{\rm enc}}=1-\frac{N}{T_H}t_{\rm
  enc}+\ldots$; the contribution of the zero-order term is annihilated by
averaging; terms of second and higher order vanish in the limit
$\hbar\to 0$ \cite{HMBH}; in the remaining first-order term $t_{\rm enc} $
cancels against the denominator whereupon the integral becomes
trivial, $\frac{1}{\Omega}\int_{-c}^c dsdu\,\e^{{\rm i}us/ \hbar}\to 2\pi\hbar/
\Omega=1/T_H$.

Higher orders in $1/N$ are due to pairs $(\gamma,\gamma')$
generated by reshuffling intra-encounter connections of loops
within an $l$-encounter with $l>2$ or within several encounters;
see Fig.~1b. In general, such reconnections arise in $v_l$
$l$-encounters, $l=2,3,\ldots$; we collect these multiplicities in
a vector $\vec v$.  The total number of encounters is
$V=\sum_{l=2}^\infty v_l$ while $L=\sum_{l=2}^\infty lv_l$ gives
the number of the encounter stretches; the total number of loops
is $L+1$.  Each encounter requires its own Poincar{\'e} section
which latter contains $l$ close-by piercings by $\gamma$;
the separations of subsequent piercings with respect to the first ones
can be expressed in terms of
coordinates $s_j,u_j,j=1,\ldots l-1$  obeying $|s_j|,|u_j|\leq c$ \cite{HMBH}. The
ergodic probability density for the $L-V$ returns is
$1/\Omega^{L-V}$. To obtain the weight function analogous to
(\ref{defw}) for the number of encounter sets with a given $\vec
v$, the ergodic density has to be integrated over all loop times
but one, and divided by the product of all encounter durations. In
addition the result has to be multiplied by the number ${\cal
N}(\vec v)$ of topological versions (``structures'') of trajectory
pairs differing by (i) the order in which encounters are visited
in $\gamma$; (ii) the relative direction of stretches in each
encounter and (iii) the intra-encounter reconnections leading from
$\gamma$ to $\gamma'$. The action difference in the pair can be
written as  $\Delta S=\sum_{\alpha,j}s_{\alpha
j}u_{\alpha j}$.  When applying the sum rule we use the exposure
time $T_{\rm exp}=\sum_{i=1}^{L+1}t_i+\sum_{\alpha=1}^Vt_{\rm
  enc}^\alpha$, as explained above. Integrating over the remaining loop
time $t_{L+1}$ (instead of the dwell time) and summing over all $\vec
v$ we obtain the complete expansion of the conductance
\begin{eqnarray}\label{tm}  \nonumber
&& {\rm Tr} (t^{\dagger} t) - \frac{N_1 N_2}{N} \\ &=& \nonumber
   \frac{N_1 N_2}{T_H}\sum_{\vec v} {\cal N}(\vec v)
   \left( \prod_{i=1}^{L+1} \int_0^{\infty} d t_i \e^{-\frac{N}{T_H} t_i}\right)
      \\ \nonumber  &&\times
      \left\langle\int\frac{ d^{L-V} u d^{L-V} s}{\Omega^{L-V}}\,
      \,\prod_{\alpha=1}^V
      \frac{\e^{- \frac{N}{T_H} t^{\alpha}_{\rm enc}+\frac{\rm i}{\hbar}
      \sum_{ j} s_{\alpha j}u_{\alpha j}}
      }{t^{\alpha}_{\rm enc}}
                     \right\rangle\\
                &=&  \frac{N_1N_2}{N}\sum_{n=1}^\infty \frac{1}{N^n}
                   \sum^{(L-V=n)}_{\vec v} (-1)^V {\cal N}(\vec v)\,.
\end{eqnarray}
Each loop is represented by one
loop-time integral as a factor and each encounter by an $(l-1)$fold
$(s,u)$-integral. The $(s,u)$-integrals are done as above, ${\rm
  e}^{-\frac{N}{T_H}t_{\rm enc}^\alpha}\to{-\frac{N}{T_H}t_{\rm enc}^\alpha} $
and $\left(\frac{1}{\Omega}\int dsdu\e^{{\rm i} su/ \hbar}\right)^{L-V}\to
T_H^{V-L}$.

In the last member of (\ref{tm}) the sum over the possible vectors
$\vec{v}$ has been split into two, one with $\vec{v}$ restricted as
$L-V=n$ and a subsequent one over the integer $n$. The partial sum
with fixed $n$ can be carried out with the help of a recursion
relation for the ${\cal N}(\vec v)$ obtained in \cite{HMBH} and reads
(see the first technical note below)
\begin{equation}\label{structuresum}
\sum_{\vec v}^{(L-V=n)} (-1)^V {\cal N}(\vec v) = (1 - 2/\beta)^n\,.
\end{equation}
Upon inserting that partial sum into (\ref{tm}) we
recover the full $1/N$-expansion (\ref{predic}) of the conductance,
now for a single device rather from an ensemble average.

\vspace{0.1cm}

The {\it transition between the orthogonal and unitary symmetry class
  in a weak magnetic field} has been discussed in
Ref.~\cite{lorenzinmagfield,Bee97, Richter}.  Saito and Nagao
\cite{Japsis} have recently accounted for a magnetic field in the
semiclassical $\tau$-expansion of the spectral form factor, up to
$\tau^3$. We here extend their ideas to the $1/N$-expansion of the
conductance. The additional action accumulated in the magnetic field
${\bf B}$ on some trajectory interval $CD$ is $\theta_{CD}=\frac{e}{c}\int_C^D {\bf A
  v}dt$ with ${\bf A}$ the vector potential. If the dynamics is
chaotic the effect of the field has been shown to be representable by
Gaussian white noise \cite{Richter} which yields the average $\left\langle
  e^{i2\theta_{CD}/\hbar}\right\rangle_G=\e^{-b\;t_{CD}}$; here $b$, scaling like $\theta_{CD}^2$, is
$\left(B/\hbar\right)^2$ times a system specific parameter. We shall see
presently that net contributions of that type come both from loops and
encounter stretches.

We consider a pair $(\gamma,\gamma')$ and
denote by $\theta_i$ the magnetic action gained on the $i$-th loop
with duration $t_i$ in $\gamma$. If that loop is passed in the
same direction by $\gamma$ and $\gamma'$ that action cancels in
$(S_\gamma-S_{\gamma'})/ \hbar$; otherwise it is doubled. Defining
a quantity $\delta_i$ equal to 0 and 1 in the first and second
case, respectively, we can write as
$\left\langle\e^{{\rm i}2\delta_i\theta_{i}/\hbar}\right\rangle_G=\e^{-b\delta_it_i}$ the
additional factor in the contribution of the pair to the
conductance associated with the loop ``$i$''. Now consider an
encounter ``$\alpha$'' and denote by $\theta_\alpha$ the magnetic
action accumulated on a single stretch of this encounter passed in
some (``positive'') direction. The magnetic action gained on all
stretches of ``$\alpha$'' in $\gamma$ will be
$\sigma_\alpha\theta_\alpha$ with $\sigma_\alpha$  the difference
between the number of the encounter stretches traversed in the
positive and negative direction. The corresponding action for the
partner $\gamma'$ will be $\sigma_\alpha'\theta_\alpha$; generally
$\sigma_\alpha'\neq \sigma_\alpha$. Averaging yields the factor
$\left\langle\e^{{\rm i}2\mu_\alpha\theta_{\alpha}/\hbar}\right\rangle_G=
\e^{-\mu_\alpha^2b\;t_{\rm enc}^{\alpha}}$ in the contribution
of the orbit pair, with the integer
$\mu_\alpha=(\sigma_\alpha-\sigma_\alpha')/2$.

We shall parametrize the magnetic field by $\xi=b T_H/N\propto (B/ \hbar)^2$
and denote by $\nu =\sum_{i=1}^{L+1} \delta_i$ the number of loops whose
sense of traversal is changed in $\gamma'$ relative to $\gamma$. Finally, we
refine our definition of the number of structures by introducing the
numbers ${\tilde{\cal N}}(\vec v, \nu ,\{\mu_{ \alpha}\})$ of topologically
different trajectory pairs with fixed $\vec v, \nu ,\{\mu_{ \alpha}\}$.  The
above result (\ref{tm}) is then reformulated for the
magnetoconductance,
\begin{eqnarray}   \label{mf}
      &&  \frac{N}{N_1N_2}\Big( {\rm Tr} (t^{\dagger} t)
- \frac{N_1N_2}{N} \Big) \\  \nonumber &=& \!\frac{1}{T_H}
\sum_{{\vec v}, \nu ,\{\mu_{ \alpha}\}}
       \!\!\!\!\! {\tilde{\cal N}}(\vec v, \nu ,\{\mu_{ \alpha}\})
   \! \prod_{i=1}^{L+1} \! \int_0^{\infty} \!\!d t_i
\e^{-(1+\xi \delta_i)t_i\frac{N}{T_H}}
    \\  \nonumber
\;\; \! \! &\times& \!\!\!\! \left\langle\!\int\!
\frac{d^{L-V} u d^{L-V} s}{ \Omega^{L-V}} \prod_{\alpha=1}^V
\frac{\e^{-\frac{N}{T_H}(1+\xi \mu^2_{\alpha})t^{\alpha}_{\rm enc}+
\frac{\rm i}{\hbar} \sum_{ j} s_{\alpha j} u_{\alpha j}}}
{ t^{\alpha}_{\rm enc}}\!\!
 \right\rangle
           \\ \nonumber
&=& \!\!\!\sum_{n=1}^\infty  \!  \frac{1}{N^{n}} \!\!\!
\sum^{(L-V = n)}_{{\vec v}, \nu ,\{\mu_{ \alpha}\}}   \!\!
        \frac{(-1)^V {\tilde{\cal N}}(\vec v, \nu ,\{\mu_{ \alpha}\}) \prod_{\alpha=1}^{V} (1+\xi
\mu^2_{\alpha}) }
{(1+\xi)^{\nu }}\,.
\end{eqnarray}
Obviously, the limiting case  $\xi \to 0$ yields the
orthogonal-class conductance (\ref{tm}) due to $\sum_{\nu
,\{\mu_{ \alpha}\}}^{(\vec{v}\,\rm
  fixed)}{\tilde{\cal N}}(\vec v, \nu,\{\mu_{ \alpha}\})={\cal
  N}(\vec{v})$; the opposite limit $\xi\to\infty$ is less transparent.
The coefficients of the resulting $1/N$-expansion for
low orders $n$ are easily obtained by checking the contributing
structures of orbit pairs for their values of $ \nu,\{\mu_{
  \alpha}\}$ and counting the multiplicities ${\tilde{\cal N}}(\vec v, \nu
,\{\mu_{ \alpha}\})$; we get
\begin{eqnarray}\label{magncond}
        {\rm Tr} (t^{\dagger} t)
&=& \frac{N_1 N_2}{N} \left[1 - \frac{1}{N}\frac{1}{(1 + \xi)}
     + \frac{1}{N^2}\frac{1}{(1 + \xi)^2}
\right.\\ \nonumber{}\\ \nonumber
&&- \left.\frac{1}{N^3}\frac{\xi^4+4\xi^3+13\xi^2+2\xi+1}{(1+\xi)^5} + \ldots \right].
\end{eqnarray}
It is seen that when $\xi\to\infty$ we do recover the unitary-class
conductance.  Order by order, our expression (\ref{magncond})
coincides with the RMT-based expansion of the integral obtained in
\cite{Weidi}. For $n>3$, it appears that our semiclassical sum
(\ref{mf}) is easier to evaluate than the corresponding term in
\cite{Weidi}. The expansion coefficients up to order $n=8$ are listed
in the second technical note below.

%%%%%%%%%%%%%%%%%%%%%%%%%%%%%%%%%%%%%%%%%%%%%%%

{\bf Technical note I:} The pairs of open trajectories of interest
here and pairs of periodic orbits with the same $\vec v$ have the same
number ${\cal N}(\vec v)$ of structures. This is a matter of
definition: we had associated in \cite{HMBH} exactly as many
structures with a single periodic orbit pair, as there are
non-equivalent trajectory pairs produced by cutting one of the orbit
loops.  We may thus use the recurrence relations for the number of
periodic orbit structures ${\cal N}(\vec{v})$ to find the sum
(\ref{structuresum}).  We need the special case incorporating
equations (44,60,30) of the online version of \cite{HMBH},
\begin{eqnarray}\label{recref}
\frac{2v_{2}}{L(\vec{v})}{\cal N}(\vec{v})&-&\frac{1}{L(\vec{v})-1}%
\sum\limits_{k\geq 2}(k+1)v_{k+1}'{\cal N}(\vec{v}\,')\nonumber\\&=&-\left( 1-%
\frac{2}{\beta }\right) {\cal N}(\vec{v}\,'')\,. \end{eqnarray}
Here $\vec{v}\,'$ (denoted $\vec{v}^{[k,2\to k+1]}$ in \cite{HMBH}) is
the vector obtained from $\vec v$ by decreasing the components $v_2$
and $v_k$ by 1 and increasing $v_{k+1}$ by 1; the vector $\,\vec{v}\,''$
($\vec{v}^{[2\to ]}$ of \cite{HMBH}) is obtained by decreasing $v_2$ by
1.  We multiply both parts of (\ref{recref}) by $(-1)^{V(\vec{v})}$
and sum over all $\vec{v}$ with fixed difference
$L(\vec{v})-V(\vec{v})=n$. In the double sum over $\vec{v},k$ we
replace $\vec{v}$ by $\vec{v}\,'$ as the summation variable and use
$%
L(\vec{v}\,')=L(\vec{v})-1,V(\vec{v}\,')=V(\vec{v})-1$; we then drop
the prime \ in $\vec{v}\,'$. Similarly, we switch in the right hand
side to $\vec{v}\,''$ as the
summation variable, taking into account $L(\vec{v}\,'')=L(
\vec{v})-2,V(\vec{v}\,'')=V(\vec{v})-1, L(\vec v\,'')-V(\vec
v\,'')=n-1$. Dropping the double
prime we obtain
\begin{eqnarray}
&&\sum\limits_{\vec{v}}^{L(\vec{v})-V(\vec{v})=n}\frac{(-1)^{V(\vec{v})}}{L(%
\vec{v})}{\cal N}(\vec{v})\left[ 2v_{2}+\sum\limits_{k\geq
2}(k+1)v_{k+1}\right]\nonumber
\\
&=&\left( 1-\frac{2}{\beta }\right) \sum\limits_{\vec{v}}^{L(\vec{v})-V(%
\vec{v})=n-1}(-1)^{V(\vec{v})}{\cal N}(\vec{v})\,.
\end{eqnarray}%
The expression in the square brackets is by definition $L(\vec{v})$
and thus cancels with the denominator. The resulting recursion
relation $S_n=(1-2/\beta)S_{n-1}$ for the sum
$S_n=\sum\limits_{\vec{v}}^{L-V=n}(-1)^{V}{\cal N}(\vec{v})$ is solved
by $S_n=S_0(1-2/\beta)^n$.  We conclude $S_n=0$ for all $n\geq 1$ in the
unitary case. For the orthogonal case we have to fix the constant
factor $S_0$.  Now for $n=1$ we have but one 2-encounter ($v_2=1$, all
other components of $\vec v$ zero) and a single structure, i.e.
${\cal N}(\vec{v})=1$ such that $S_1=-1$ and thus $S_0=1$.

{\bf Technical note II:} We here extend the $1/N$ expansion
(\ref{magncond}) of the conductance in the presence of a magnetic
field. Defining the expansion coefficients as
\begin{equation}
{\rm Tr}(t^\dagger t)=\frac{N_1N_2}{N}\sum_{n=0}^\infty\frac
{C_n(\xi)}{N^{n}}\;,\label{eightordr}
\end{equation} and introducing the abbreviations $q_k(\xi)\equiv 1+k^2 \xi,\quad
p(\xi)\equiv 1/(1+\xi)=q_1^{-1}$ we have, by numerically determining the numbers
${\tilde{\cal N}}(\vec v, \nu ,\{\mu_{ \alpha}\})$,
\begin{eqnarray}
C_0&=&1,\quad C_1=-p,\quad C_2=p^{2},\nonumber \\
C_3&=&-p+q_3p^{4}-q_2^2p^{5},\nonumber\\ C_4&=&p^{2}-p^{4}-4q_3p^{5}+5q_2^2p^{6},\nonumber \\
C_5&=&-p+q_3p^4-q_2^2p^5+8q_5p^6-24q_2q_4p^7\nonumber\\&&-12q_3^2p^7+49q_2^2q_3p^8-21q_2^4p^9,
\nonumber\\
C_6&=&p^2-p^4-4q_3p^5+5q_2^2p^6+4p^6+16q_3p^7
\nonumber\\&&-48q_5p^7
+168q_2q_4p^8-21q_2^2p^8+84q_3^2p^8
\nonumber\\&&-392q_2^2q_3p^9
+189q_2^4p^{10},\nonumber\\
C_7&=&
-p+ p^{ 4}  q_3
  -p^{ 5}  q_2^2
+ 8p^{ 6}  q_5
  -12p^{ 7}  q_3^2
   -24p^{ 7}  q_2  q_4
\nonumber\\
&&+180p^{ 8}  q_7
  +49p^{ 8}  q_2^2 q_3 -21p^{ 9}  q_2^4
  -608p^{ 9}  q_3  q_5
 \nonumber\\
&& -276p^{ 9}  q_4^2
  -720p^{ 9}  q_2  q_6 + 1728p^{10}  q_2^2 q_5
 \nonumber\\
&& + 2952p^{10}  q_2  q_3  q_4  +
   464p^{10}  q_3^3 -3240p^{11}  q_2^3 q_4
 \nonumber\\&&
   -4440p^{11}  q_2^2 q_3^2
 + 5445p^{12}  q_2^4 q_3
   -1485p^{13}  q_2^6\;,\nonumber\\
C_8&=&
      p^{  2} +  4p^{  6}   -p^{  4} -36p^{  8}
  + 216p^{ 10}  q_2^2
   -21p^{  8}  q_2^2\nonumber\\%
&&
 -1485p^{ 12}  q_2^4
     +   5p^{  6}  q_2^2
  + 189p^{ 10}  q_2^4\nonumber\\%
&& +19305p^{ 14}  q_2^6 -160p^{  9}  q_3
  +16p^{  7}  q_3
 \nonumber\\%
 && +2880p^{ 11}  q_2^2 q_3
    -4p^{  5}  q_3-392p^{  9}  q_2^2 q_3
    \nonumber\\%
&&-65340p^{ 13}  q_2^4 q_3
 -1080p^{ 10}  q_2 q_4
  + 168p^{  8}  q_2 q_4\nonumber\\%
&&+35640p^{ 12}  q_2^3 q_4
  -624p^{ 10}  q_3^2
  +  84p^{  8}  q_3^2\nonumber\\
&& + 48840p^{ 12}q_2^2 q_3^2  + 288p^{  9}  q_5
   -48p^{  7}  q_5
\nonumber\\%
&&-17280p^{ 11}  q_2^2 q_5 -29520p^{ 11}  q_2 q_3 q_4
 -4640p^{ 11}  q_3^3
\nonumber\\%
&& + 6480p^{ 10}  q_2 q_6
 + 2484p^{ 10}  q_4^2
 + 5472p^{ 10}  q_3 q_5\nonumber\\
&& -1440p^{  9}  q_7\,.\nonumber
\end{eqnarray}
A recursive treatment of all orders in the expansion (\ref{eightordr})
is called for.

%%%%%%%%%%%%%%%%%%%%%%%%%%%%%%%%%%%%%%%%%%%%%%

We are indebted to Dmitry Savin for useful discussions and the
Sonderforschungsbereich SFB/TR12 of the Deutsche
Forschungsgemeinschaft for financial support.


\begin{thebibliography}{notitle}
\bibitem{footRS} In Ref.~\cite{Richter} the self-crossings
  of a trajectory were allowed to approach the cavity
  openings which means in our language that half of each encounter
  stretch could stick out into the leads; this is equivalent to
  replacing $(T-2t_{\rm enc})^2$ by $(T-t_{\rm enc})^2$ in
  (\ref{defw}); in addition, the increase of the survival probability
  due to the self-encounter was not taken into account. The two
  differences to our procedure cancel in their effect on the
  lowest off-diagonal term $G^{(1)}$.
\end{thebibliography}
\end{document}